# Acceleration of protons to high energies by an ultra-intense femtosecond laser pulse

Jarosław Domański , Jan Badziak , Sławomir Jabłoński
Institute of Plasma Physics and Laser Microfusion, 01 – 497 Warsaw, Poland

**ABSTRACT**

The paper reports the results of two-dimensional particle-in-cell simulations of proton beam acceleration at the interactions of a 130-fs laser pulse of intensity from the range of $10^{21} – 10^{23}$ W/cm$^2$, predicted for the Extreme Light Infrastructure (ELI) lasers currently built in Europe, with a thin hydrocarbon (CH) target. A special attention is paid to the effect of the laser pulse intensity and polarization (linear - LP, circular - CP) as well as the target thickness on the proton energy spectrum, the proton beam spatial distribution and the proton pulse shape and intensity. It is shown that for the highest, ultra-relativistic intensities (~ $10^{23}$ W/cm$^2$) the effect of laser polarization on the proton beam parameters is relatively weak and for both polarizations quasi-monoenergetic proton beams of the mean proton energy ~ 2 GeV and $\delta E/E \approx 0.3$ for LP and $\delta E/E \approx 0.2$ for CP are generated from the 0.1-μm CH target. At short distances from the irradiated target (< 50 um), the proton pulse is very short (< 20 fs), and the proton beam intensities reach extremely high values > $10^{21}$ W/cm$^2$, which are much higher than those attainable in conventional accelerators. Such proton beams can open the door for new areas of research in high energy-density physics and nuclear physics as well as can also prove useful for applications in materials research e.g. as a tool for high-resolution proton radiography.

## 1. INTRODUCTION

Laser-driven ion accelerators have a potential to be applied in various branches of science, technology and medicine and are considered to be a compact, flexible and cost-effective alternative to the conventional ion sources based on RF-driven accelerators [1-3]. However, to produce ion beams of parameters required for applications petawatt (PW) or multi-PW short-pulse lasers have to be used as the accelerator drivers and mechanisms and properties of the ion acceleration process using such lasers should be deeply and fully understood. The mechanisms of ion acceleration and parameters of generated ion beams depend significantly on both the laser beam and irradiated target parameters, in particular on the laser beam intensity and the thicknesses and composition of the target. The ions can be accelerated by several laser-induced mechanisms such as: the target normal sheath acceleration (TNSA) [2-5], the radiation pressure acceleration (RPA)[2,3,6-8] (also known as the skin-layer ponderomotive acceleration – SLPA [9-11]), the laser break-out afterburner (BOA) [2,3,12,13], the collisionless electrostatic shock acceleration (CESA) [2,3,14,15] or the laser-induced cavity pressure acceleration (LICPA) [16,17]. Since the dominant acceleration mechanism is determined by the laser-target interaction conditions, the laser beam and target parameters should be carefully selected to produce the ion beams of characteristics required for a particular application. Moreover, the ion acceleration mechanisms must be well identified and controlled.

Extreme Light Infrastructure (ELI) is a currently implemented large-scale European project that uses cutting-edge laser technologies to build multi-PW lasers generating femtosecond pulses of ultra-relativistic intensities ~ $10^{22} – 10^{23}$ W/cm$^2$ [18,19]. Parameters of these laser pulses seem to be sufficient to generate ion beams required for various applications; however, the studies of ion acceleration in the ultra-relativistic intensity regime are in a very initial stage [6,13,20,21] and need to be intensely developed. In particular, it concerns the acceleration of protons that at such laser intensities can reach sub-GeV and GeV energies which are required for applications in nuclear physics and can also be useful in materials research or cancer therapy. This paper presents selected results of our numerical studies of proton generation from a thin hydrocarbon (CH) target irradiated by a 130 fs laser of ultra-high intensity predicted for the ELI lasers and ranging from $10^{21}$ W/cm$^2$ to $10^{23}$ W/cm$^2$. The effect of the laser pulse intensity and polarization as well as the target thickness on characteristics of the generated proton beam is investigated with the use of relativistic two-dimensional (2D) particle-in-cell (PIC) code. In particular, it is shown that the ELI lasers are capable of producing GeV proton beams of very short duration and extremely high intensities unattainable in conventional RF-driven accelerators operating presently.

## 2. RESULTS AND DISCUSION

The investigations of the interaction of a femtosecond laser pulse with the CH target and ion acceleration were performed using fully electromagnetic relativistic 2D PIC code elaborated at the Institute of Plasma Physics and Laser Microfusion, Warsaw. The correctness of the code was verified by a comparison of the results obtained using the code with the ones from measurements presented in [11,22] as well as with the results of 3D simulations of carbon ion acceleration carried out by Sgattoni et al. [23]. The numerical simulations were performed for the CH target of the areal mass density $\sigma_t$ ranging from 0.06 to 0.8 mg/cm$^3$ and the transverse dimension equal to 12 µm. Molecular density of the target corresponded to solid-state density and was equal to $4.86*10^{22}$ molecules/cm$^3$ and the target components (C, H) were fully ionized. A pre-plasma layer of 0.25 µm thickness and the density shape described by an exponential function was placed in front of the target. The laser pulse shape in time and space (along the y-axis) was described by a super-Gaussian function of the power index equal to 6. The laser beam width (FWHM-Full Width at Half Maximum) was assumed to be 8 µm and the laser pulse length (FWHM) and wavelength were equal to 130 fs and 800 nm, respectively. The simulations were performed in the s, y space of dimensions 160*32 µm$^2$ and the number of macroparticles was assumed to be $5*10^6$ for thinner targets ($\sigma_t \leq 0.2$ mg/cm$^3$) and $9*10^6$ for thicker targets ($\sigma_t > 0.2$ mg/cm$^3$).

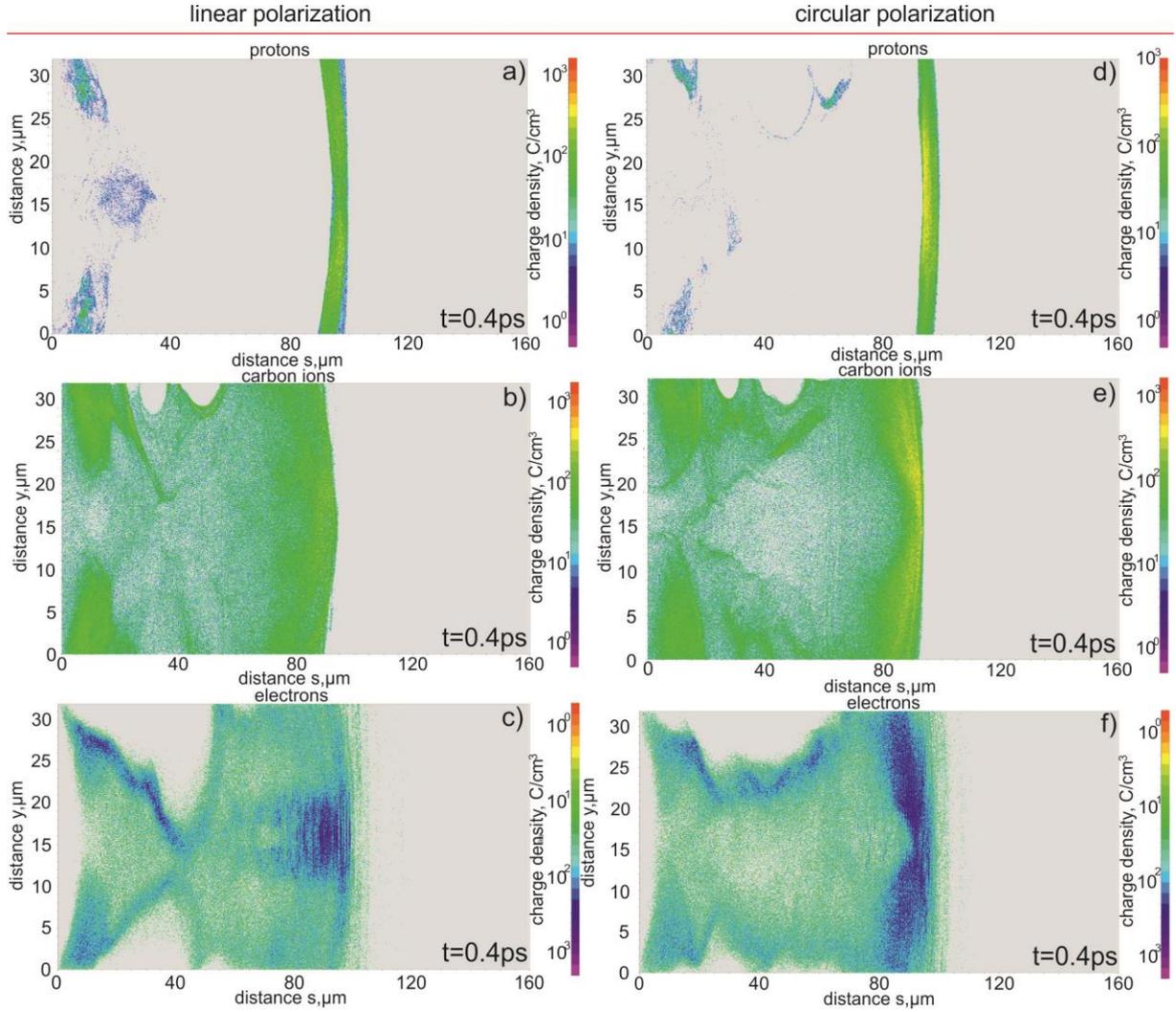

Figure 1. 2D spatial distributions of charge density of protons (a, d), carbon ions (b, e) and electrons (c, f). The target was irradiated by the circulaly-LP (a, b, c) or linearly-CP (d, e, f) polarized laser pulses of intensity equal to $10^{23}$ W/cm$^2$ for CP and $2*10^{23}$ W/cm$^2$ for LP. t=0.4ps. $\sigma_t$ =0.06mg/cm$^2$.

Figure 1 presents a 2D spatial distribution of charge density of protons (a, d), carbon ions (b, e) and electrons(c, f). The target was irradiated by the linearly-LP (a, b, c) and circularly-CP (d, e, f) polarized laser pulses of the intensity equal to $10^{23}$ W/cm$^2$ for CP and $2*10^{23}$ W/cm$^2$ for LP (the laser energy fluencies was approximately the same for both types of polarization). The presented situation corresponds to the final stage of ion acceleration (the simulation time is equal to 0.4 ps). The weak influence of laser polarization on the acceleration process can be observed. This effect could be explained by the dominance of the RPA mechanism in the ultra-relativistic intensity regime investigated here. In this mechanism, the mean energy of accelerated ions depends on the laser energy fluence and the target areal density which were identical for both laser polarization in the cases considered. An important feature of proton acceleration in the presented cases is a small velocity dispersion of the accelerated protons that move with the relativistic velocities in the form of high-density proton (plasma) block. It can also be seen that a fairly large part of carbon ions move with the velocities close to the proton velocities.

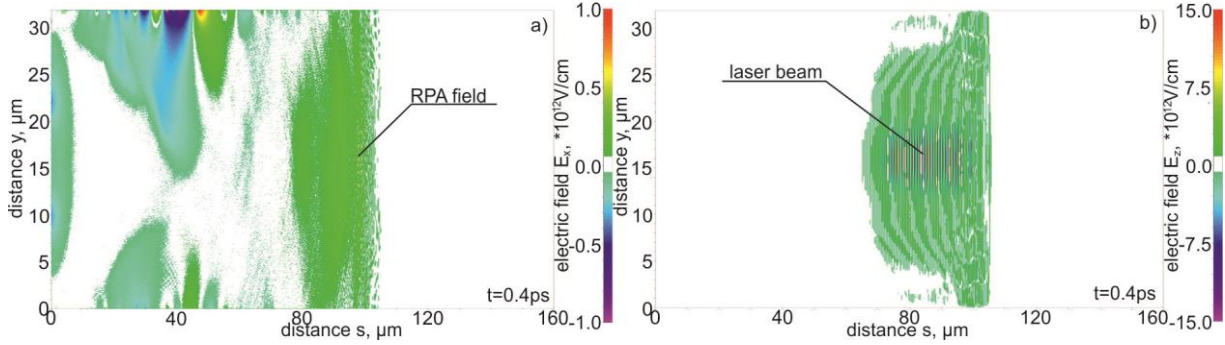

Figure 2. 2D spatial distributions of the electric field component parallel (a) and perpendicular (b) to the laser beam propagation direction for LP and for the late stage of acceleration. $I_L=2*10^{23}$ W/cm$^2$, t=0.4 ps, $\sigma_t$ =0.06mg/cm$^2$.

Figure 2 shows 2D spatial distributions of electric field components parallel (a) and perpendicular (b) to the laser beam propagation direction for the linear polarization case. The parallel component of the electric field is directly responsible for ion and electron acceleration. The perpendicular one represents part of the laser field modified due to the interaction with plasma. It is visible that the RPA mechanism is the dominant one, and that protons and a significant part of carbon ions are accelerated by this mechanism.

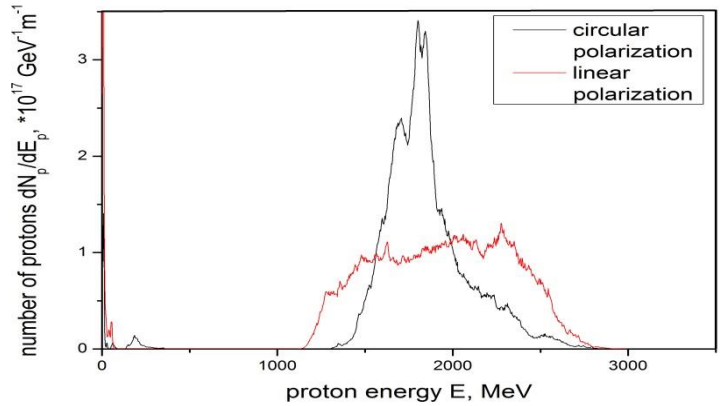

Figure 3. The energy spectra of protons accelerated forward from the hydrocarbon target irradiated by the laser pulse of linear or circular polarization. t=0.4ps, $I_L=10^{23}$ W/cm$^2$ for CP and $I_L=2*10^{23}$ W/cm$^2$ for LP, $\sigma_t$ =0.06mg/cm$^2$.

The following figure (Fig. 3) presents the energy spectra of the accelerated protons for the late stage of acceleration (t = 0.4 ps). As opposed to the cases of lower laser intensities ($\leq 10^{21}$ W/cm$^2$) investigated in [24,25], the shapes of these spectra are rather similar to the Gaussian distribution. For this reason, the parameters appropriate to describe them are the mean ion energy and the standard ion energy deviation. The mean energy of protons for both types of investigated polarization is equal to 1.8 GeV; however, the proton energy dispersion is smaller for the circular polarization than for the linear one. The standard energy deviation expressed in the mean energy units is equal to $dE_p/\bar{E}_p = 0.19$ for CP and $dE_p/\bar{E}_p = 0.3$ for LP. The broader energy spectrum for LP could be explained by the fluctuation of the ponderomotive force accelerating protons observed in the case of LP and not observed for CP.

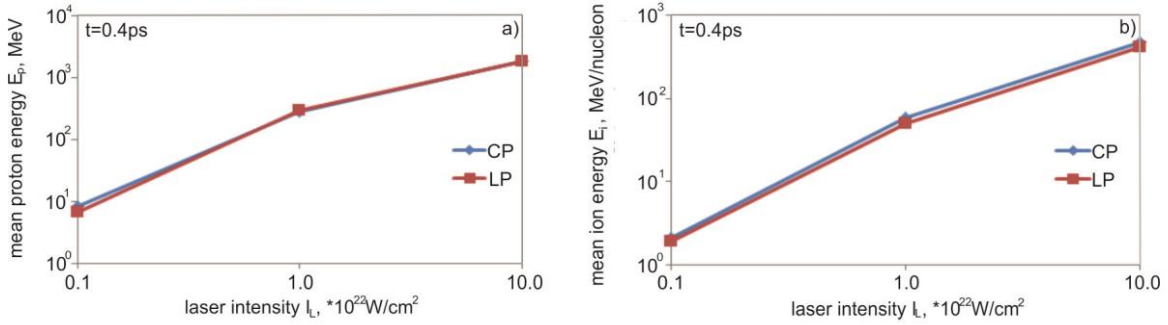

Figure 4. The mean energies of protons and carbon ions for both types of laser polarization as a function of laser beam intensity. t=0.4ps, $\sigma_t$ =0.06mg/cm$^2$.

The mean energies of protons and carbon ions for both types of polarization as a function of laser beam intensity are shown in Fig. 4 (the laser energy fluence is the same for both polarizations and the laser intensities on the horizontal axis correspond to the circular polarization). In general, the influence of laser polarization on the proton and carbon ion energies is weak as it was stated above. However, this influence depends on laser intensity and is more significant for lower laser intensities. It can be explained by the fact that the lower laser intensity, the greater the contribution of the TNSA mechanism to the acceleration process, which mechanism is strongly influenced by the laser polarization.

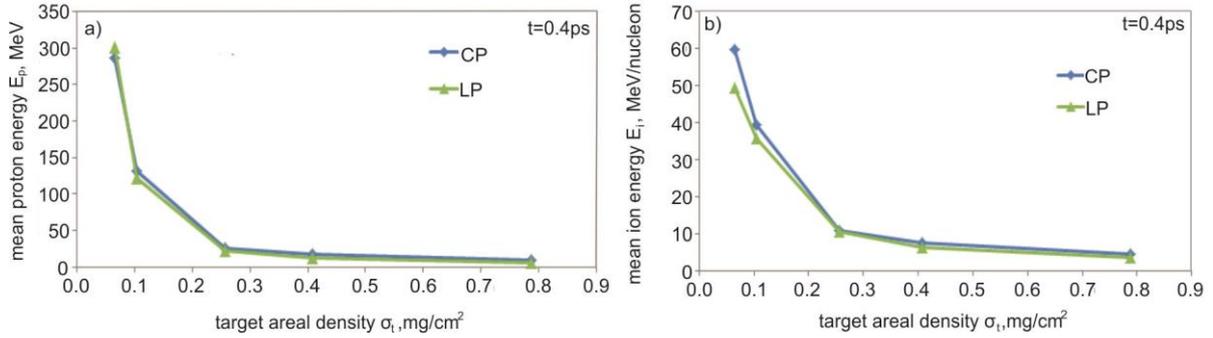

Figure 5. The quantitative values of the mean energy for protons and carbon ions as a function of the areal target density $\sigma_t$. t=0.4ps, $I_L=10^{22}$ W/cm$^2$ for CP and $I_L=2*10^{22}$ W/cm$^2$ for LP.

The dependences of the mean energies for protons and carbon ions on the areal target density $\sigma_t$ are shown in Fig. 5. The mean energy of both protons and carbon ions is approximately inversely proportional to the thickness of the target and the highest values are achieved for the thinnest target.

The characteristics of ion beams such as the beam intensity and the shape and duration of the ion pulse are important from the point of view of various potential applications of laser-accelerated ions, and in the applications such as the ion fast ignition of inertial fusion, the production of high-energy-density matter or some nuclear physics experiments, they play a key role [1-3]. Fig. 6 presents the temporal distribution of the intensity of the proton beam driven by the laser pulse of linear or circular polarization recorded 40 μm behind the rear surface of the target. Similarly, as for the characteristics of proton beam discussed previously, the weak influence of laser polarization on the temporal shape and the peak intensity of the proton beam can be seen. For both types of investigated polarizations, the proton pulses of duration about 11 fs and peak intensity of $8\times10^{21}$ W/cm$^2$ are produced. These proton pulses are by several orders of magnitude shorter and their intensities are much higher than those produced in conventional RF-driven accelerators. The dependence of the peak proton pulse intensity on laser beam intensity for both types of laser beam polarization is shown in Fig. 7. For the low-intensity range ($10^{21} – 10^{22}$ W/cm$^2$) the proton pulse intensity grows very quickly with the laser intensity (faster than linearly) and for the ultra-relativistic laser intensity range (> $10^{22}$ W/cm$^2$) it increases approximately proportionally to the laser intensity and reaches extremely high values $\geq 10^{21}$ W/cm$^2$.

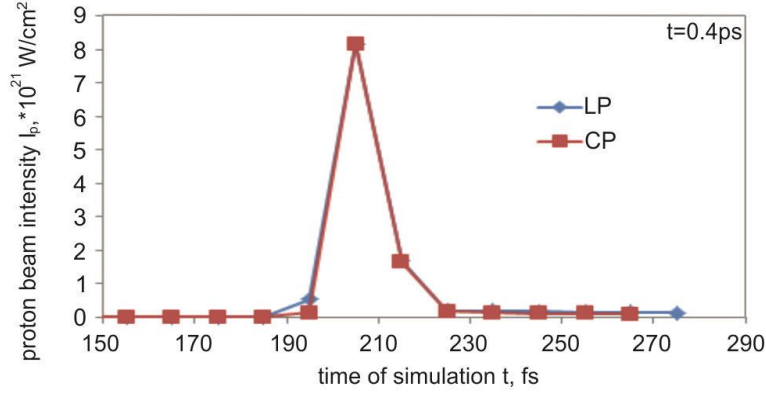

Figure 6. The temporal distribution of intensity of the proton beam driven by the laser pulse of linear or circular polarization. The distributions were recorded 40 μm behind the rear surface of the target. $I_L=10^{23}$ W/cm$^2$ for CP and $I_L=2*10^{23}$ W/cm$^2$ for LP, $\sigma_t$ =0.06mg/cm$^2$.

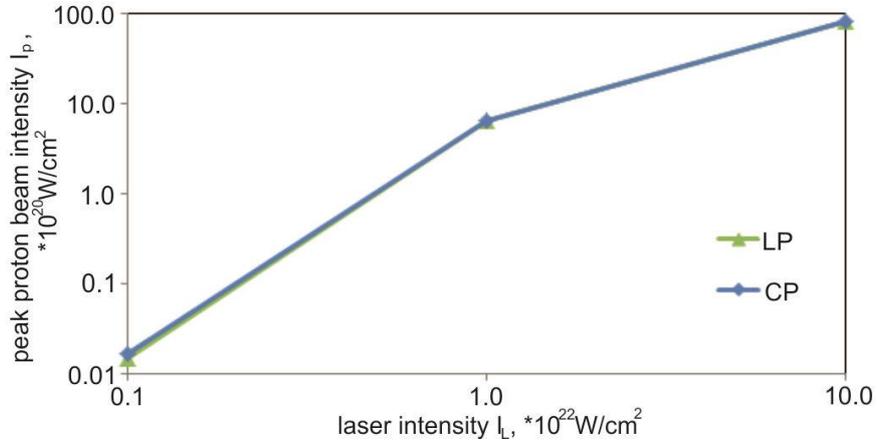

Figure 7. The peak proton beam intensity as a function of laser beam intensity for both types of laser beam polarization. The proton beam intensity distributions were recorded 40 μm behind the rear surface of the target. $\sigma_t$ =0.06mg/cm$^2$.

Although most of accelerated protons and carbon ions move together with electrons in the form of the quasi-neutral plasma block, the electron distribution in plasma does not cover exactly the ion distributions (Fig. 1) and locally the net electric charge in the plasma is different from zero. The inhomogeneous distribution of the charge in the plasma leads to the generation of electric current the total net value $J = (J_p + J_C) - J_e$ of which can be fairly high ($J_e$, $J_p$ and $J_C$ are the electron current, the proton current and the carbon ion current, respectively). The total current J changes in time and is unevenly distributed in space as demonstrated in Figs. 8 and 9. Fig. 8 presents the temporal run of the total current (integrated over the whole transverse size of the simulation box) at the distance of 40 um from the target rear surface while Fig. 9 shows the transverse distribution of the current density at this distance at t = 0.4 ps. It can be observed that at the time longer (at least several times) than the laser pulse duration the total current is dominated by fast electrons (the current is negative) and reaches the value of several MA. This strong flux of fast electrons escaping from the target is the generator of an intense electromagnetic emission (usually in the THz domain of frequency) and, moreover, it is the cause of the creation of a net positive charge of the target (an electric neutrality of the target is disturbed by the loss of negative charge carried by the fast electron flux). The neutralization current flowing through the target and the target holder can be the source of a strong electromagnetic pulse (EMP), usually in the GHz domain, which propagates into the interaction chamber and may be harmful for diagnostic equipment and other electronic devices [26,27]. Since, in general, the higher the power of the laser pulse interacting with the target, the stronger is EMP, the problem of EMP generated in the laser-target interaction is especially important for big laser facilities like ELI and is currently a hot topic intensely studied. Our simulations, in particular the results presented in Figs. 8 and 9, show briefly the way the fundamental source of EMP is created in the sub-ps time scale when the ion beam is formed and accelerated due to the laser-target interaction. More detailed analysis of this issue is beyond the scope of the present paper and will be done in our other papers.

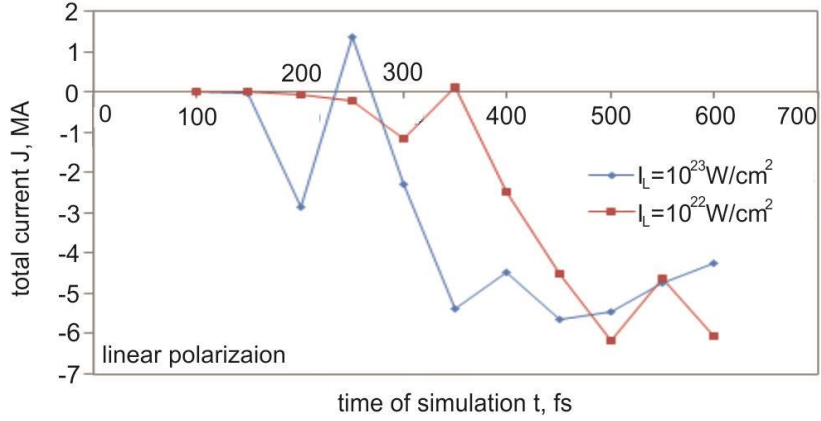

Figure 8. The temporal run of the net total current recorded at the distance of 40 μm from the target rear surface. $\sigma_t$ =0.06mg/cm$^2$, linear polarization.

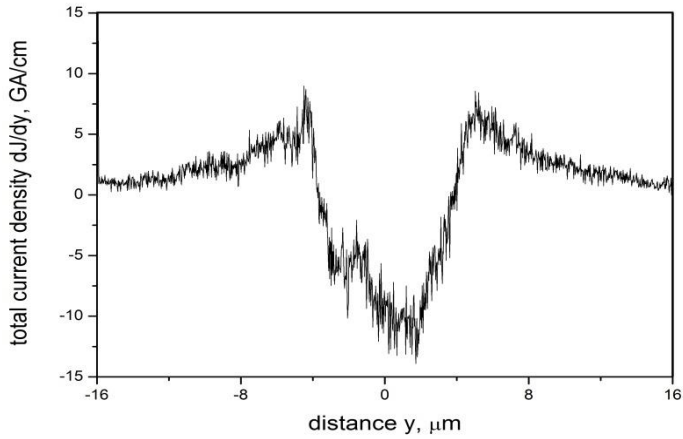

Figure 9. The transverse distribution of the net total current density recorded at the distance of 40 μm from the target rear surface at t=0.4ps. $I_L$=10$^{23}$ W/cm$^2$, linear polarization, . $\sigma_t$ =0.06mg/cm$^2$.

## 3. CONCLUSIONS

In conclusion, the properties of proton beam generation at the interaction of fs laser pulse of ultra-relativistic intensity with a thin, sub-micrometer hydrocarbon target have been investigated with the use of 2D PIC code. It has been found that for the laser intensities above 10$^{22}$ W/cm$^2$ parameters of the generated proton beam weakly depend on the laser beam polarization and a dominant mechanism of proton/ion acceleration is RPA. At laser intensity ~ 10$^{23}$ W/cm$^2$ a relativistic, quasi-monoenergetic proton beam of the mean proton energy ~ 2 GeV is produced. At short distances from the irradiated target (< 50 um), the proton pulse is very short (< 20 fs), and the proton beam intensities reach extremely high values > 10$^{21}$ W/cm$^2$, which are much higher than those attainable in conventional RF-driven accelerators operating presently. Such proton beams can open the door for new areas of research in high energy-density physics and nuclear physics as well as can also be useful for inertial confinement fusion studies or for materials research e.g. as a tool for high-resolution proton radiography.

## ACKNOWLEDGMENTS

This work was supported in part by the National Centre for Science (NCN), Poland under the Grant No. 2014/14/M/ST7/00024. The simulations were carried out with the support of the Interdisciplinary Center for Mathematical and Computational Modelling (ICM), University of Warsaw under grant no. G57-20.

# REFERENCES

[1] Ledingham, K.W.D. and Galster, W., "Laser-driven particle and photon beams and some applications", *New J. Phys.* **12**, 045005 (2010).

[2] Daido, H., Nishiuchi, M. and Pirozhkov, A.S., "Review of laser driven ion sources and their applications", *Rep. Prog. Phys.* **75**, 056401 (2012).

[3] Macchi, A., Borghesi, M. and Passoni, M., "Ion acceleration by superintense laser-plasma interaction", *Rev. Mod. Phys.* **85**, 751 (2013).

[4] Wilks, S.C., Langdon, A.B., Cowan, T.E., Roth, M., Singh, M., Hatchett, S., Key, M.H., Pennington, D., MacKinnon, A. and Snavely, R.A., "Energetic proton generation in ultra-intense laser-solid interactions", *Phys. Plasmas* **8**, 542 (2001).

[5] Borghesi, M., Fuchs, J., Bulanov, S.V., MacKinnon, A.J., Patel, P.K. and Roth, M., "Fast ion generation by high-intensity laser irradiation of solid targets and applications", *Fusion Sci. Technol.* **49**, 412 (2006).

[6] Esirkepov, T., Borghesi, M., Bulanov, S.V., Mourou, G. and Tajima, T., "Highly efficient relativistic-ion generation in the laser-piston regime", *Phys. Rev. Lett.* **92**, 175003 (2004).

[7] Macchi, A., Cattani, F., Liseykina, T.V. and Cornalti, F., "Laser acceleration of ion bunches at the front surface of over-dense plasmas", *Phys. Rev. Lett.* **94**, 165003 (2005).

[8] Robinson, A.P.L., Zepf, M., Kar, S., Evans, R.G. and Bellei, C., "Radiation pressure acceleration of thin foil with circular polarized laser pulse", *New J. Phys.* **10**, 013021 (2008).

[9] Badziak, J., Jabłoński, S. and Głowacz, S., "Generation of highly collimated high-current ion beams by skin-layer laser-plasma interaction at relativistic intensities", *Appl. Phys. Lett.* **89**, 061504 (2006).

[10] Badziak, J., Głowacz, S., Hara, H., Jabłoński, S. and Wołowski, J., "Studies on laser-driven generation of fast high-density plasma blocks for fast ignition", *Laser Part. Beams* **24**, 249 (2006).

[11] Badziak, J., Jabłoński, S., Parys, P., Rosiński, M., Wołowski, J., Szydłowski, A., Antici, P., Fuchs, J. and Mancic, A., "Ultraintense proton beams from laser-induced skin-layer ponderomotive acceleration", *J. Appl. Phys.* **104**, 063310 (2008).

[12] Yin, L., Albright, B.J., Hegelich, B.M. and Fernandez, J.C., "GeV laserion acceleration from ultrathin targets: the laser breakout afterburner", *Laser Part. Beams* **24**, 291-298 (2006).

[13] Yin, L., Albright, B.J., Hegelich, B.M., Browers, K.J., Flippo, K.A., Kwan, T.J.T. and Fernandez, J.C., "Monoenergetic and GeV ion acceleration from the laser breakout afterburner using ultrathin targets", *Phys. Plasmas* **14**, 056706 (2007).

[14] Denavit, J., "Absorption of high-intensity subpicosecond lasers on solid density targets", *Phys. Rev. Lett.* **69**, 3052 (1992).

[15] Silva, L.O., Marti, M., Davies, J.R. and Fonseca, R.A., "Proton shock acceleration in laser-plasma interactions", *Phys. Rev. Lett.* **92**, 015002 (2004).

[16] Badziak, J., Borodziuk, S., Pisarczyk, T., Chodukowski, T., Krokusy, E., Masek, J., Skala, J., Ullschmied, J. and Rhee, Y.-J., "Highly efficient acceleration and collimation of high-density plasma using laser-induced cavity pressure", *Appl. Phys. Lett.* **96**, 251502 (2010).

[17] Badziak, J., Jabłoński, S., Pisarczyk, T., Rączka, P., Krokusky, E., Liska, R., Kucharik, M., Chodukowski, T., Kalinowska, Z., Parys, P., Rosiński, M., Borodziuk, S. and Ullschmied, J., „ Highly efficient accelerator of dense matter using laser-induced cavity pressure acceleration", *Phys. Plasmas* **19**, 053105 (2012).

[18] www.eli-laser.eu

[19] Danson, C., Hillier, D., Hopps, N. and Neely, D., "Petawatt class lasers worldwide", *High Power Laser Sci. Eng.* **3**, e3 (2015).

[20] Zheng, F.L., Wang, H.Y., Yan, X.Q., Tajina, T., Yu M.Y. and He X.T., "Sub-TeV proton beam generation by ultra-intense irradiation of foil-and-gas target", *Phys. Plasmas* **19**, 023111 (2012).

[21] Xu, Y., Wang, J., Qi, X., Li, M., Xing, Y., Yang, L. and Zhu, W., "Plasma black acceleration via double targets driven by an ultraintense circularly polarized laser pulse", *Phys. Plasmas* **24**, 033108 (2017).

[22] Badziak, J., Antici, P., Fuchs, J., Jabłoński, S., Mancic, A., Parys, P., Rosiński, M., Suchańska, R., Szydłowski, A. and Wołowski, J., "Laser-induced generation of ultraintense proton beams for high energy-density science", *AIP Conf. Proc.* **1024**, 63-77 (2008).

[23] Sgattoni, A., Sinigardi, S. and Macchi, A., "High energy gain in three-dimensional simulations of light sail acceleration", *Appl. Phys. Lett.* **105**, 084105 (2014).

[24] Domański, J., Badziak, J. and Jabłoński, S., "Numerical studies of petawatt laser-driven proton generation from two-species targets using a two-dimensional particle-in-cell code", *J. Instrum.* **11**, C04009 (2016).

[25] Domański, J., Badziak, J. and Jabłoński, S., "Enhanced efficiency of femtosecond laser-driven proton generation from a two-species target with heavy atoms", *Laser Part. Beams* **34**, 294-298 (2016).

[26] Dubois, J.-L., Lubrano-Lavaderci, F., Raffestin, D., Ribolzi, J., Gazave, J., Compant La Fontaine, A., d'Humieres, E., Hulin, S., Nicolai, Ph., Poye, A. and Tikhonchuk, V.T., "Target charging in short-pulse-plasma experiments", *Phys. Rev. E* **89**, 013102 (2014).

[27] Poye, A., Dubois, J.-L., Lubrano-Lavaderci, F., d'Humieres, E., Bardon, M., Hulin, S., Bailly-Grandvaux, M., Ribolzi, J., Raffestin, D., Santos, J.J., Nicolai, Ph. and Tikhonchuk V., " Dynamic model of target charging by short laser pulse interactions", *Phys. Rev. E* **92**, 043107 (2015).